\documentstyle[12pt]{article}
\begin{document}
\begin{center}
{\Large\bf Consistent Batalin--Fradkin quantization\\
\bigskip
{\Large\bf of Infinitely Reducible First Class Constraints}}\\
\bigskip
\vskip 0.7cm
{\large Stefano {Bellucci}\footnote{E-mail: bellucci@lnf.infn.it}
and Anton {Galajinsky}\footnote{E-mail: agalajin@lnf.infn.it}}
\vskip 0.4cm
{\it INFN--Laboratori Nazionali di Frascati, C.P. 13, 
00044 Frascati, Italia}\\

\end{center}

\vskip 0.5cm

\begin{abstract}
We reconsider the problem of BRST quantization of a mechanics with
infinitely reducible first class constraints. Following an earlier recipe
[Phys. Lett. B{\bf 381} 105 (1996)], the original phase space is extended
by purely auxiliary variables, the constraint set in the enlarged space being
first stage of reducibility. The BRST charge involving only a finite number 
of ghost variables is explicitly constructed. 
\end{abstract}

\vspace{0.4cm}


\twocolumn

The problem of infinitely reducible first class constraints originated 
from the superstring theory where a fully satisfactory covariant quantization
seems to be an unsolved problem yet. Taking a simpler mechanics analogue 
in four dimensions these look like  
\begin{equation}\label{s}
p^2=0, {(p_{\theta} \sigma^n p_n)}_{\dot\alpha}=0,
{(\sigma^n p_{\bar\theta} p_n)}_\alpha=0,
\end{equation}
where $(p_n,p_{\theta\alpha},p_{\bar\theta \dot\alpha})$ are momenta
conjugate to the variables parametrizing a conventional $R^{4|4}$ 
superspace $(x^n,\theta^\alpha,{\bar\theta}^{\dot\alpha})$ and 
${\sigma^n}_{\alpha\dot\alpha}$ are the Pauli matrices. Owing to the 
null vector $p_n$ entering the problem, only half of the fermionic 
constraints is linearly independent. In particular, the identity
\begin{equation}\label{pr}
{(p_{\theta} \sigma^n p_n)}_{\dot\alpha} {Z_1}^{\dot\alpha \alpha}
+{Z_1}^\alpha p^2\equiv0,
\end{equation}
where ${Z_1}^{\dot\alpha \alpha}={({\tilde\sigma}^n p_n)}^{\dot\alpha 
\alpha}, {Z_1}^\beta={p_\theta}^\beta$, holds. 
On the constraint surface not all of the functions 
${Z_1}^{\dot\alpha \alpha}$ prove to be independent 
\begin{equation}\label{pr}
{Z_1}^{\dot\alpha \alpha}{Z_2}_{\alpha\dot\beta}\approx0,\quad
{Z_2}_{\alpha\dot\beta}={(\sigma^n p_n)}_{\alpha\dot\beta}.
\end{equation}
Apparently, this process can be continued, the system at hand being  
infinite stage of reducibility~\cite{b}. It is worth mentioning that, 
although the correct counting of degrees of freedom can be achieved in 
the course of BRST quantization by making use of Euler's 
regularization~\cite{k}, the expression for the 
BRST charge involves an infinite ghost tower~\cite{tow} and, hence, 
looks formal. 

A recipe how to supplement infinitely reducible first class constraints 
up to a constraint system of finite stage of reducibility has been proposed
recently~\cite{d}. It suffices to extend the original phase space
by purely auxiliary variables 
$(\Lambda^n,p_{\Lambda m})$,$(\chi^\alpha,p_{\chi\alpha})$,
$({\bar\chi}_{\dot\alpha},\\
{p_{\bar\chi}}^{\dot\alpha})$, 
with $\Lambda$ being a real boson and $(\chi,\bar\chi)$ a pair of 
complex conjugate fermions.
These are required to satisfy reducible constraints like
those in Eq. (\ref{s}) (one can check that the number and the class of 
the constraints are just enough to suppress dynamics in 
the sector~\cite{d})
\begin{equation}\label{add1} 
p_{\chi\alpha}=0,\quad {(\chi \sigma^n \Lambda_n)}_{\dot\alpha}=0, 
\end{equation}
\begin{equation}\label{add2} 
p_{\bar\chi \dot\alpha}=0, \quad {( \sigma^n \bar\chi \Lambda_n)}_{\alpha}=0, 
\end{equation}
\begin{equation}\label{add3}
p_{\Lambda n}=0,\quad \Lambda^2=0,\quad 1-\Lambda p=0.  
\end{equation}
Beautifully enough, in the extended phase space the reducibility of the 
original constraints (\ref{s}) can be compensated by that coming from 
the sector of additional variables to put the fermionic constraints in
the irreducible form 
\begin{equation}\label{fin1}
{\bar\Phi}_{\dot\alpha} \equiv {(p_\theta \sigma^n p_n+p_{\chi}\sigma^n 
\Lambda_n)}_{\dot\alpha}=0,
\end{equation}
\begin{equation}\label{fin2}
\Phi_\alpha \equiv {(p_n \sigma^n p_{\bar\theta}+\Lambda_n \sigma^n 
p_{\bar\chi})}_\alpha=0,
\end{equation}
\begin{equation}\label{fin3}
{\bar\Psi}_{\dot\alpha} \equiv {(\chi \sigma^n \Lambda_n+p_{\chi}\sigma^n 
p_n)}_{\dot\alpha}=0, 
\end{equation}
\begin{equation}\label{fin4}
\Psi_\alpha \equiv {(\Lambda_n \sigma^n \bar\chi+p_n \sigma^n 
p_{\bar\chi})}_\alpha=0,
\end{equation}
while in the bosonic sector one has
\begin{equation}\label{fin5}
p^2=0,
\end{equation}
\begin{equation}\label{fin6}
{\tilde p}_{\Lambda m} \equiv p_{\Lambda m}-(p_\Lambda \Lambda)p_m-
(p_\Lambda p)\Lambda_m=0,
\end{equation}
\begin{equation}\label{fin7}
p_\Lambda p=0, \Lambda^2=0, p_\Lambda \Lambda=0, 1-\Lambda p=0.
\end{equation}
The equivalence to the initial constraint set seems to be more transparent
if one makes use of the identity
\begin{eqnarray}\label{id1}
&& {p_\chi}^\alpha=-{\textstyle{\frac {1}{2\Lambda p}}}p^2 {p_\theta}^\alpha-
{\textstyle{\frac {1}{2\Lambda p}}}\Lambda^2 \chi^\alpha-\nonumber\\
&&{\textstyle{\frac {1}{2\Lambda p}}}{\bar\Phi}_{\dot\alpha} 
{({\tilde\sigma}^m p_m)}^{\dot\alpha}-{\textstyle{\frac {1}{2\Lambda p}}}
{\bar\Psi}_{\dot\alpha}{({\tilde\sigma}^m 
\Lambda_m)}^{\dot\alpha \alpha},\nonumber\\
\end{eqnarray}
and its complex conjugate.
In the new basis the constraints (\ref{fin1}),(\ref{fin2}),(\ref{fin5}),
(\ref{fin6}) are first class, whereas Eqs.~(\ref{fin3}),(\ref{fin4}) and
(\ref{fin7}) involve second class ones. In order to 
explicitly decouple ${\tilde p_\Lambda}^n=0$ from the second 
class fermionic constraints it suffices to 
redefine them like ${\tilde p_\Lambda}^n=0 \rightarrow 
{\tilde p_\Lambda}^n-{\textstyle{\frac 12}}\chi\sigma^n{\tilde\sigma}^m 
p_\chi p_m-{\textstyle{\frac 12}}p_{\bar\chi}{\tilde\sigma}^m \sigma^n 
\bar\chi p_m\\
=0$. As the Dirac bracket associated with the second class 
constraints is introduced, this seems to be 
inessential here.

Residual reducibility proves to fall in the bosonic sector.
Due to the identities (in what follows the symbol $\approx$ denotes an 
equality up to a linear combination of {\it second class} constraints)
\begin{equation}\label{toz}
{\tilde p}_\Lambda \Lambda \approx 0, \qquad {\tilde p}_\Lambda p \approx 0,
\end{equation}
there are only two linearly independent components 
entering Eq.~(\ref{fin6}), {\it the system in the extended phase space being 
first stage of reducibility}. 

It is the purpose of the present Brief Report to explicitly construct the 
BRST charge associated with the constraint set~(\ref{fin1})--(\ref{fin7}), 
thus giving an efficient way to cure the infinite ghost tower problem 
intrinsic to the original system~(\ref{s}).

According to the general recipe~\cite{b} the nilpotency 
equation to determine the BRST charge should be solved 
under the Dirac bra-\\
cket associated to the second class 
constraints. Evaluated in specific coordinate sectors this 
reads (only the brackets to be used 
below are explicitly given here) 
\begin{eqnarray}\label{brin}
&&\{\chi^\alpha,p_{\chi\beta}\}={\textstyle{\frac 12}}
{\delta^\alpha}_\beta-{\textstyle{\frac {2}{\Delta}}} 
\Lambda p {{(\sigma_{nm})}_\beta}^\alpha \Lambda^n p^m,\nonumber\\
&&\{\chi^\alpha,\chi^\beta\}={\textstyle{\frac {2}{\Delta}}}p^2
{(\sigma_{nm})}^{\alpha\beta} \Lambda^n p^m,\nonumber\\ 
&& \{p_{\chi\alpha},p_{\chi\beta}\}=
{\textstyle{\frac {2}{\Delta}}}\Lambda^2
{(\sigma_{nm})}_{\alpha\beta} \Lambda^n p^m;
\end{eqnarray}
\begin{eqnarray}
&&\{\Lambda^n,p_{\Lambda m}\}={\delta^n}_m+
{\textstyle{\frac {2}{\Delta}}} p^2 \Lambda^n \Lambda_m+\nonumber\\
&&\qquad {\textstyle{\frac {2}{\Delta}}}\Lambda^2 p^n p_m-
{\textstyle{\frac {2}{\Delta}}}\Lambda p (p^n\Lambda_m
+\Lambda^n p_m),\nonumber\\
&&\{p_{\Lambda n},p_{\Lambda m}\}=
{\textstyle{\frac {2}{\Delta}}}p^2
(\Lambda_n p_{\Lambda m}-\Lambda_m p_{\Lambda n})+\nonumber\\
&&\qquad \qquad {\textstyle{\frac {2}{\Delta}}}p p_\Lambda (p_n\Lambda_m-
p_m\Lambda_n)+\nonumber\\
&&\qquad \qquad {\textstyle{\frac {2}{\Delta}}}p\Lambda (p_{\Lambda n}p_m-
p_{\Lambda m}p_n)-\nonumber\\
&&\qquad \qquad {\textstyle{\frac {i}{\Delta}}}
(\chi^2-{\bar\chi}^2)\epsilon_{nmkl}\Lambda^k p^l,\nonumber\\
&&\{\Lambda^n,\Lambda^m \}=0;
\end{eqnarray}
\begin{eqnarray}
&&\{\theta^\alpha,p_{\theta\beta}\}={\delta^\alpha}_\beta, \quad
\{\theta^\alpha,\theta^\beta\}=0,\nonumber\\  
&& \{p_{\theta\alpha},p_{\theta\beta}\}=0; \quad \{p_n,p_m\}=0,
\end{eqnarray}
plus complex conjugate expressions for the pairs
$(\bar\chi,p_{\bar\chi})$,$(\bar\theta,p_{\bar\theta})$.

In the cross sectors the only non 
vanishing brackets are (in what follows we will not need the 
explicit form of the brackets involving $x^n$--variable, these
are omitted here)
\begin{eqnarray}
&& \{p_{\Lambda n},\chi^\alpha\}={\textstyle{\frac {1}{\Delta}}}p^2
(\Lambda_n\chi^\alpha+{(\chi \sigma_n {\tilde\sigma}^k 
\Lambda_k)}^\alpha)
\nonumber\\
&&\qquad \qquad +{\textstyle{\frac {1}{\Delta}}}p_n
{(\chi \sigma^k \Lambda_k {\tilde\sigma}^m p_m)}^\alpha-\nonumber\\
&& \qquad \qquad {\textstyle{\frac {1}{\Delta}}}\Lambda p {(\chi \sigma_n
{\tilde\sigma}^k p_k)}^\alpha,
\end{eqnarray}
\begin{eqnarray}
&&\{p_{\Lambda n},p_{\chi\alpha}\}=
{\textstyle{\frac {1}{\Delta}}}\Lambda^2
(p_n\chi_\alpha+{(\chi \sigma_n {\tilde\sigma}^k p_k)}_\alpha)\nonumber\\
&&\qquad \qquad+{\textstyle{\frac {1}{\Delta}}}\Lambda_n
{(\chi \sigma^k p_k {\tilde\sigma}^m \Lambda_m)}_\alpha-\nonumber\\
&&\qquad \qquad {\textstyle{\frac {1}{\Delta}}}\Lambda p 
{(\chi \sigma_n {\tilde\sigma}^k \Lambda_k)}_\alpha,
\end{eqnarray}
plus complex conjugates.

Given the Dirac bracket, the algebra of the first class constraints is easy 
to evaluate
\begin{eqnarray}\label{alg}
&& \{ {\tilde p}_{\Lambda n},{\tilde p}_{\Lambda m}\}\approx
{U_{nm}}^k{\tilde p}_{\Lambda k}+
U_{nm}p^2,\nonumber\\
&&\{ {\tilde p}_{\Lambda n},\Phi_\alpha \}\approx{U_{n\alpha}}^\beta 
\Phi_\beta+ 
U_{n\alpha}p^2,\nonumber\\
&&\{ {\tilde p}_{\Lambda n},{\bar\Phi}_{\dot\alpha} \}
\approx{U_{n\dot\alpha}}^{\dot\beta} {\bar\Phi}_{\dot\beta}
+U_{n\dot\alpha}p^2,
\end{eqnarray}
with all other brackets vanishing. The structure functions entering
Eq.~(\ref{alg}) are given by
\begin{eqnarray}
&& {U_{nm}}^k={\textstyle{\frac {2}{\Delta}}}
((\Lambda_n p^2-p_n){\delta_m}^k-\nonumber\\
&& \qquad \qquad (\Lambda_m p^2-p_m){\delta_n}^k),\nonumber\\
&& U_{nm}={\textstyle{\frac {i}{\Delta}}}(p_\chi \chi-p_{\bar\chi}\bar\chi)
\epsilon_{nmkl} \Lambda^k p^l,\nonumber\\
&&{U_{n\alpha}}^\beta={\textstyle{\frac 12}}
{{(\sigma_n {\tilde\sigma}^k p_k)}_\alpha}^\beta+
{\textstyle{\frac {1}{\Delta}}}\Lambda_n p^2 {\delta_\alpha}^\beta+\nonumber\\
&&\qquad \qquad{\textstyle{\frac {1}{\Delta}}}(\Lambda_n p^2-p_n) 
{{(\Lambda^k \sigma_k {\tilde\sigma}^l p_l)}_\alpha}^\beta,\nonumber\\
&& U_{n\alpha}={\textstyle{\frac 12}}{(\sigma_n p_{\bar\theta})}_\alpha-
{\textstyle{\frac {1}{\Delta}}}\Lambda_n{(p^k \sigma_k 
p_{\bar\theta})}_\alpha+\nonumber\\
&&{\textstyle{\frac {1}{\Delta}}}(\Lambda_n p^2-p_n) {(\Lambda^k \sigma_k 
p_{\bar\theta})}_\alpha,
\end{eqnarray}
and ${U_{n \dot\alpha}}^{\dot\beta}={({U_{n\alpha}}^\beta)}^{*}$,
$U_{n \dot\alpha}={(U_{n\alpha})}^{*}$. Worth noting also is the 
orthogonality of the structure functions obtained to the vectors 
$p_n, \Lambda^n$ which holds on the second class constraints 
surface.

Having evaluated the structure functions, we are now in a position to
construct the BRST charge. Associated with the first class constraints
(\ref{fin1}),(\ref{fin2}),(\ref{fin5}),(\ref{fin6}) are the primary ghosts 
(minimal sector) $(C^{\dot\alpha},{\bar{\cal P}}_{\dot\alpha})$, 
$(C^\alpha,{\bar{\cal P}}_\alpha)$,\\
$(C,\bar{\cal P})$,$(C^n,{\bar{\cal P}}_n)$. These have the standard properties
\begin{eqnarray} 
&& \epsilon (C^A)=\epsilon ({\bar{\cal P}}^A)=\epsilon_A +1, \nonumber\\
&& gh (C^A)=-gh ({\bar{\cal P}}^A)=1.
\end{eqnarray}
To compensate the overcounting in the sector $(C^n,{\bar{\cal P}}_n)$
(only two components entering Eq.~(\ref{fin6}) are linearly independent)
one further introduces the secondary ghosts~\cite{b} $(C^1,{\bar{\cal P}}^1)$,
$(C^2,{\bar{\cal P}}^2)$, these obeying 
\begin{eqnarray} 
&& \epsilon (C^{1,2})=\epsilon ({\bar{\cal P}}^{1,2})=0, \nonumber\\
&& gh (C^{1,2})=-gh ({\bar{\cal P}}^{1,2})=2.
\end{eqnarray}
The nilpotency equation on the BRST charge
\begin{equation}\label{nilp}
\{ \Omega_{min},\Omega_{min} \} \approx0,
\end{equation}
should then be solved under the boundary condition
\begin{eqnarray}\label{bound}
&&\Omega_{min}=\Phi_\alpha C^\alpha +{\bar\Phi}_{\dot\alpha} C^{\dot\alpha}
+{\tilde p}_{\Lambda n} C^n+p^2 C\nonumber\\
&&\quad \quad +{\bar{\cal P}}_n \Lambda^n C^1
+{\bar{\cal P}}_n p^n C^2+\dots,
\end{eqnarray}
which, through~(\ref{nilp}), automatically generates both the algebra
(\ref{alg}) and the identities~(\ref{toz}). 
 
Calculating the contribution of the boundary terms into the 
equation~(\ref{nilp})
\begin{eqnarray}
&& \{ \Omega_{min},\Omega_{min} \} \approx 
2{\bar{\cal P}}_m \{\Lambda^m,{\tilde p}_{\Lambda n} \} 
C^1 C^n- \nonumber\\ 
&& \quad \quad 2({U_{n\alpha}}^\beta \Phi_\beta+U_{n\alpha} 
p^2)C^\alpha C^n- \nonumber \\
&& \quad \quad 2({U_{n\dot\alpha}}^{\dot\beta} 
{\bar\Phi}_{\dot\beta}+U_{n\dot\alpha} p^2) C^{\dot\alpha} 
C^n-\nonumber\\
&& \quad \quad ({U_{nm}}^k {\tilde p}_{\Lambda k}+U_{nm} p^2) 
C^m C^n+\dots,\nonumber\\
\end{eqnarray}
one can partially clarify the structure of the terms lacking in
Eq.~(\ref{bound}). In particular, extending the ansatz~(\ref{bound}) 
by the three new contributions 
\begin{eqnarray}
&& {\textstyle{\frac 12}}{\bar{\cal P}}_k {{\tilde U}_{nm}}^k C^m C^n
+{\bar{\cal P}}_\alpha {U_{n\beta}}^\alpha C^\beta C^n+\nonumber\\
&& {\bar{\cal P}}_{\dot\alpha} {U_{n\dot\beta}}^{\dot\alpha}
C^{\dot\beta} C^n, 
\end{eqnarray}
with
\begin{eqnarray}
&& {{\tilde U}_{nm}}^k={U_{nm}}^k-{\textstyle{\frac {2}{\Delta}}} p^k 
(\Lambda_n p_m-\Lambda_m p_n),\nonumber\\  
&& {{\tilde U}_{nm}}^k \Lambda^m\approx {\textstyle{\frac {2}{\Delta}}} 
\{ \Lambda^k,p_{\Lambda n} \}, {{\tilde U}_{nm}}^k p^m\approx 0,\nonumber\\  
\end{eqnarray}
one can get rid of the first term (which is a manifestation
of reducibility of the constrains) and those involving
${\tilde p}_\Lambda,\Phi,\bar\Phi$
\begin{eqnarray}\label{corr}
&& \{ \Omega_{min},\Omega_{min} \}\approx 
-U_{nm}p^2 C^m C^n-\nonumber\\
&&\quad \quad 2U_{n\alpha} p^2 C^\alpha C^n-2U_{n\dot\alpha} 
p^2 C^{\dot\alpha} C^n-\nonumber\\
&&\quad 2{\bar{\cal P}}_\alpha {U_{n\gamma}}^\alpha 
{U_{m\beta}}^\gamma C^m C^n C^\beta-\nonumber\\
&&\quad 2{\bar{\cal P}}_{\dot\alpha} {U_{n\dot\gamma}}^{\dot\alpha} 
{U_{m\dot\beta}}^{\dot\gamma} C^m C^n C^{\dot\beta}+\dots.   
\end{eqnarray}
In order to verify Eq.~(\ref{corr}) a number of Jacobi identities
associated to the constraint algebra~(\ref{alg}) should be used. 
These are omitted here.

It is instructive then to give the explicit form of the terms 
quadratic in the structure functions which enter Eq. (\ref{corr})
($ {U_{m{\dot\alpha}}}^{\dot\beta} {U_{n{\dot\beta}}}^{\dot\gamma}-
{U_{n{\dot\alpha}}}^{\dot\beta} {U_{m{\dot\beta}}}^{\dot\gamma}$ 
is obtained by complex conjugation) 
\begin{eqnarray}
&& {U_{m\alpha}}^\beta {U_{n\beta}}^\gamma-{U_{n\alpha}}^\beta 
{U_{m\beta}}^\gamma=\{ {{(\sigma_{nm})}_\alpha}^\beta+ \nonumber\\
&& {\textstyle{\frac {1}{\Delta}}}(\Lambda_n p_m-\Lambda_m p_n)
{{(\Lambda_l \sigma^l {\tilde\sigma}^k p_k)}_\alpha}^\gamma+\nonumber\\
&&{\textstyle{\frac {1}{\Delta}}} 
\Lambda_m{{(\sigma_n {\tilde\sigma}^k p_k)}_\alpha}^\gamma-
{\textstyle{\frac {1}{\Delta}}} \Lambda_n{{(\sigma_m 
{\tilde\sigma}^k p_k)}_\alpha}^\gamma-\nonumber\\
&&{\textstyle{\frac {1}{\Delta}}} (\Lambda_m p^2 -p_m){{(\sigma_n 
{\tilde\sigma}^k \Lambda_k)}_\alpha}^\gamma+\nonumber \\
&&{\textstyle{\frac {1}{\Delta}}} 
(\Lambda_n p^2 -p_n){{(\sigma_m {\tilde\sigma}^k 
\Lambda_k)}_\alpha}^\gamma+\nonumber\\
&& {\textstyle{\frac {1}{\Delta}}}(\Lambda_n p_m-\Lambda_m p_n)
{\delta_\alpha}^{\gamma}
 \}p^2 \equiv {\Pi_{mn\alpha}}^\gamma p^2.\nonumber\\
\end{eqnarray}

Being factors of $p^2$ these suggest a further amendment 
\begin{eqnarray}
&&\bar{\cal P} U_{n\alpha} C^\alpha C^n+\bar{\cal P} U_{n\dot\alpha} 
C^{\dot\alpha} C^n+\nonumber\\
&&{\textstyle{\frac 12}} \bar{\cal P} U_{nm} C^m C^n-
{\textstyle{\frac 12}} \bar{\cal P} {\bar{\cal P}}_\alpha  
{\Pi_{nm\beta}}^\alpha C^m C^n C^\beta-\nonumber\\
&&{\textstyle{\frac 12}} \bar{\cal P} {\bar{\cal P}}_{\dot\alpha}  
{\Pi_{nm\dot\beta}}^{\dot\alpha} C^m C^n C^{\dot\beta}.
\end{eqnarray}
After tedious calculations with the extensive use of Jacobi identities
one can verify that the complete BRST charge 
\begin{eqnarray}\label{BRST}
&& \Omega_{min}=\Phi_\alpha C^\alpha +{\bar\Phi}_{\dot\alpha} 
C^{\dot\alpha}+
{\tilde p}_{\Lambda n} C^n+\nonumber\\
&&p^2 C+{\bar{\cal P}}_n \Lambda^n C^1+{\bar{\cal P}}_n p^n C^2+ \nonumber\\
&& {\textstyle{\frac 12}}{\bar{\cal P}}_k {{\tilde U}_{nm}}^k C^m C^n
+{\bar{\cal P}}_\alpha {U_{n\beta}}^\alpha C^\beta C^n+\nonumber\\
&&{\bar{\cal P}}_{\dot\alpha} {U_{n\dot\beta}}^{\dot\alpha} 
C^{\dot\beta} C^n+\bar{\cal P} U_{n\alpha} C^\alpha C^n+\nonumber\\
&&\bar{\cal P} U_{n\dot\alpha} C^{\dot\alpha} C^n+
{\textstyle{\frac 12}} \bar{\cal P} U_{nm} C^m C^n- \nonumber \\
&& {\textstyle{\frac 12}} \bar{\cal P} {\bar{\cal P}}_\alpha  
{\Pi_{nm\beta}}^\alpha C^m C^n C^\beta-\nonumber\\
&&{\textstyle{\frac 12}} \bar{\cal P} 
{\bar{\cal P}}_{\dot\alpha}  {\Pi_{nm\dot\beta}}^{\dot\alpha} 
C^m C^n C^{\dot\beta},
\end{eqnarray}
is nilpotent. Beautifully enough, only a finite number of ghost generations
proved to be needed in the extended phase space.

Finally, it is worth mentioning that a formal consideration of the present 
paper can be directly applied to specific models. In particular,
the superparticle due to Siegel~\cite{s1}, after a proper
Hamiltonian treatment, leads precisely to Eq. (1) we started with. The
latter theory has been previously considered in the alternative harmonic
superspace approach~\cite{nis} . This makes 
use of Lorentz harmonics~\cite{nis} in order to extract linearly 
independent components 
from the fermionic constraints~(\ref{s}) in a covariant way. 
Having obtained a system of rank two, our result here is in perfect
agreement with that of Ref.~\cite{nis}. The present formulation, however, has 
the advantage that all the variables involved obey the standard 
spin--statistics relations. Furthermore, the scheme outlined in this 
article proves to admit a Lagrangian formulation~\cite{d},\cite{bell}, 
the latter seems to be problematic in the approach~\cite{nis}.

Another interesting approach to be mentioned is that by Diaz and 
Zanelli~\cite{dz} who improved an earlier (noncovariant) quantization 
proposal by Kallosh~\cite{kal2}
(see also related work~\cite{bell1}). The infinite proliferation of 
ghosts has been truncated there by
imposing appropriate conditions on the ghosts variables, the latter involving
specific (covariant) projectors. In this respect, the possibility to
truncate the infinite ghost tower at the second step following the
approach by Diaz and Zanelli seems to be an interesting further
point to confirm our technique. This and other questions related
to possible applications to superparticle, superstring will be considered
in a forthcoming publication~\cite{bell}.

\end{document}